\documentclass[pra,aps,twocolumn,epsf,amsfonts]{revtex4}

\usepackage{graphicx}% Include figure files
\usepackage{epsfig}
\def\qed{\leavevmode\unskip\penalty9999 \hbox{}\nobreak\hfill
     \quad\hbox{\leavevmode  \hbox to.77778em{%
               \hfil\vrule   \vbox to.675em%
               {\hrule width.6em\vfil\hrule}\vrule\hfil}}
     \par\vskip3pt}

\def\ibb #1{\leavevmode\hbox{\kern.3em\vrule
     height 1.5ex depth -.1ex width .4pt\kern-.3em\rm#1}}

\def\ket #1{\vert #1\rangle}
\def\braket #1#2{\langle #1 \vert #2\rangle}
\def\ketbra #1#2{\vert #1\rangle \! \langle #2\vert}

\newcommand\beq{\begin{equation}}
\newcommand\eeq{\end{equation}}
\newcommand\bea{\begin{eqnarray}}
\newcommand\eea{\end{eqnarray}}

\begin{document}

\title{Interpolation of recurrence and hashing entanglement distillation protocols}
\author{Karl Gerd H. Vollbrecht and Frank Verstraete}
\address{Max-Planck-Institut f\"ur
Quantenoptik, Hans-Kopfermann-Str. 1, D-85748 Garching, Germany}
\date{\today}

\begin{abstract}
We construct new entanglement distillation protocols by
interpolating between the recurrence and hashing protocols. This
leads to asymptotic two-way distillation protocols, resulting in
an improvement of the distillation rate for all mixed Bell
diagonal entangled states, even for the ones with very high
fidelity. We also present a method how entanglement-assisted
distillation protocols can be converted into
non-entanglement-assisted protocols with the same yield.
\end{abstract}


\maketitle

\section{Introduction}

One of the big achievements in entanglement  theory has been to prove that it is
possible to get around the effects of decoherence using an amount of resources
that only scales linearly with the number of pure resources one would need.
Entanglement distillation and quantum error correction enables to distribute
maximally entangled states between different parties in the present of a noisy
environment and imperfect quantum channels. This basic task is one of the
requirements, if we want  to apply the new applications provided by Quantum
information theory in a realistic unperfect situation.

Entangled states, in particular pure maximally entangled states, are used as a
resource in many protocols in quantum information theory and {\it entanglement of
distillation } is the most reasonable measure for entanglement respecting this
resource character. Given $n$ copies of a bipartite state $\rho$, where the both
subsystems are respectively owned by two distinct physicists called  {\it Alice }
and {\it Bob}, the task is to transform these states into $m$ copies of maximally
entangled states. To fulfill this task, Alice and Bob are allowed to make any
sequence of local operations on her/his site and they can coordinate their efforts
via a classical channel. The set of operations accessible in this way is called
LOCC (local operations and classical communication). The best possible  rate $m/n$
they could archive (in the asymptotic limit of many copies) defines the
distillable entanglement. The latter quantifies in some sense the amount of useful
entanglement contained in the state $\rho$. Furthermore, one can think of a
situation, where Alice and Bob have already a (maybe infinite) pool of
predistilled maximally entangled states. The goal is now to enlarge the number of
maximally entangled states in their pool at the end of the protocol, but in
between maximally entangled states can be abused for establishing non local
operation. Such a protocol is called an {\it entanglement assisted protocol}.

Generically, starting with a finite number $n$ of copies, it is
never possible to generate maximally entangled states by means of
LOCC operations. Purely maximally entangled states can only be
achieved by an  asymptotic protocol, i.e., in the limit of the
number of copies going to infinity.

The construction of such asymptotic distillation protocols leading to a non zero rate
is a nasty task and essentially all known protocols are only improved versions of
the {\it hashing/breeding} distillation protocol
presented in \cite{BennettPRL,BennettPRA}, which is adapted to
Bell diagonal states of a two qubit system. This protocol was recently generalized to arbitrary states
in arbitrary dimensions \cite{Winter}.

The hashing/breeding protocol is a so called {\it one way
distillation protocol}, what means, that the classical information
is only sent in one direction, e.g. from Alice to Bob, but not in
the other way. The great advantage of such kind of protocols is,
that they are directly related to quantum error correction codes
\cite{BennettPRA}. But it is well known, that one way distillation
protocols are not optimal \cite{BennettPRA}.

One easy way to upgrade a one way distillation protocol to a two
way distillation protocol, is to add a prefixed two way operation,
like a pre-selection of states based on a measurement outcome,
acting respectively only on a few copies. The states produced this
way are used as new input for the one way distillation protocol.
The prototype of such a protocol is the recurrence protocol, where
two copies of a states are mapped by a two way operation to a
higher entangled states, which is distilled by the (one way)
hashing protocol. Much effort has been spent to optimize or modify
such kind of recurrence methods
\cite{Smolin,AlberJex,Ekert,Frank}. The problem that all this
protocols share, is that the two way communication part is not
asymptotic,
 resulting in an improvement of the distillation rates only for states with relatively low entanglement.

The present paper is devoted to develop the first  asymptotical two way
 distillation protocol.
  The following results are obtained:

\begin{itemize}
  \item We provide an entanglement assisted  two way distillation protocol
   for Bell diagonal two qubit states.

  \item We show this protocol to give a strictly positive improvement
  for all Bell diagonal states, expect for the low rank cases, where the hashing protocol
  was already known to be optimal.

  \item  We show how to further improve the rates and calculate some examples
  on Werner states.

  \item We show how to transform this protocols into distillation protocols, that did not need
  any pre-distilled entanglement.

  \item We hope to give a better understanding of the breeding protocol and its relation
  to the recurrence protocol.
\end{itemize}

\section{Preliminaries}
The new distillation protocol  we want to introduce will be
applied to Bell diagonal states on a two qubit system. In
particular we will discuss the case of Werner states. Let us start with a short introduction to these kinds of states and  briefly
recall same basics about the fundamental operations used in the
hashing/breeding protocol.

 For a two
qubit Hilbert  space ${\cal H}=\mathbb{C}^2\otimes\mathbb{C}^2$ there
exists a basis of maximally entangled states given by the so
called four Bell states:

\bea
\psi_{00}=\frac{1}{\sqrt{2}}(\ket{00}+\ket{11})\\
\psi_{01}=\frac{1}{\sqrt{2}}(\ket{00}-\ket{11})\\
\psi_{10}=\frac{1}{\sqrt{2}}(\ket{10}+\ket{01})\\
\psi_{11}=\frac{1}{\sqrt{2}}(\ket{10}-\ket{01}). \eea The
projectors onto these Bell states will be denoted by
$P_{ij}=\ketbra{\psi_{ij}}{\psi_{ij}}$. We will in the following
consider states of the form
\begin{equation}\label{rholambda}
\rho_{\lambda} = \sum_{l,k=0}^{1} \lambda_{kl} P_{kl},
\end{equation}
which are  called Bell diagonal states. These states are paramerized by the four
eigenvalues $\lambda:=\{\lambda_{00},\lambda_{01},\lambda_{10},\lambda_{11}\}$ .
Bell diagonal states play an important role in entanglement theory
\cite{BennettPRA,BennettPRL,HHHred}, especially in distillation theory. Every
entangled two qubit state can be transformed to an entangled Bell diagonal states
by means of LOCC operations \footnote{An optimal single-copy procedure to do this
with probability 1 has been found in \cite{VDD01}, while the optimal probabilistic
protocol is given in \cite{VV03}.}. So the understanding of distillation
protocols for Bell diagonal states gives a deep insight in the distillation of
arbitrary states.

A Bell diagonal states is separable, if and only if none of its
eigenvalues exceeds $\frac{1}{2}$. It is well known, that every
entangled Bell diagonal state can be distilled by a combination of
the recurrence and the breeding/hashing protocol
\cite{BennettPRA}, but the distillation rates archived so far,
seem to be far away from an optimum and are much to low to be used
as a good lower bound for distillable entanglement.

A Bell diagonal state is called a Werner state \cite{Werner, iso}, if all its
eigenvalues $\lambda_{ij}$ except one, e.g.  $\lambda_{00}$ are equal. Such a
state can  be written as $$\rho_f=f P_{00}+\frac{1-f}{3}({\bf1}-P_{00})$$ The free
parameter $f \equiv\lambda_{00} $ is called fidelity. A Werner state is entangled,
if and only if $f>0.5$.

As in the case of the known hashing, breeding or recurrence
protocol, our protocol will completely stay in the framework of
Bell diagonal states, i.e. it will only consist of operations,
that map many copies of Bell states to many copies of Bell states.
In detail, we will use only bilateral CNOT operations and local
measurements.

An essential ingredient for any distillation protocol acting on
two qubit systems is the CNOT operation. The CNOT operation is
defined by \beq C\ket{i,j}=\ket{i ,(i+j)}, \eeq where the first
tensor factor is called the {\it source} and the second the {\it
target}. The addition $(i+j)$ should be read as modulo 2. It is
readily verified that a {\it bilateral} CNOT operation (BCS),
where both parties in a bipartite system apply CNOT operations
locally on a tensor product of two Bell states, acts as
\begin{equation}\label{BCS}
(C\otimes C ) |\psi_{ij}\rangle\otimes |\psi_{kl}\rangle =
|\psi_{i,j+ l}\rangle\otimes|\psi_{k+ i,l}\rangle,
\end{equation}
here the first tensor product on the l.h.s. in (\ref{BCS})
corresponds to the Alice$|$Bob split, whereas the others
correspond to the source$|$target split. A bilateral CNOT
operation maps two copies of a Bell state again to two copies of a
Bell state.

In the framework of Bell states, i.e. maximally entangled states,
every outcome of a local measurement on Alice's or Bobs side is
completely random. Therefore, local measurement makes only sense,
if Alice and Bob compare their result. If Alice and Bob make both
a measurement in the $\ket{0},\ket{1}$ basis, the difference of
their outcomes will tell them whether it was one of the Bell
states $\psi_{00},\psi_{01}$ or one of the states
$\psi_{10},\psi_{11}$. So Alice and Bob can "measure" together the
"$i$" of an unknown Bell state $\psi_{ij}$. In the same way they
can measure "$j$" using the $\ket{0+1},\ket{0-1}$ basis. We will
refer to such local measurements as {\it local Bell measurements}.

\section{Known Protocols}
Our new protocol is based one the known hashing/breeding protocol and the recurrence protocol. To explain the new protocol
it is extremely  helpful to give a briefly sketch of these to methods. Later on we will explain our new results. Anybody
feeling quite familiar with these protocols can immediately move to section \ref{new}.
\subsection{The recurrence method}
The recurrence method takes as input two copies
of a Bell diagonal state $\rho_\lambda$, one called the target
state, the other the source state.
 The idea is to apply a bilateral CNOT operation  on two copies of
the state $\rho_\lambda$ and then making a local Bell measurement on the target state.
 The source states is kept whenever the
measurement outcomes of Alice and Bob coincide, otherwise it is
discarded. The overlap of the resulting state with the maximally
entangled states increases, if the original overlap was larger
than $1/2$, which can always be obtained for entangled Bell
diagonal states. The protocol was originally  introduced only for
Werner states, where at the end of the protocol the resulting
state is mapped to a Werner state again.
 By iterating this method one can produce from a entangled
Werner state  (Bell diagonal states) a state that is arbitrarily close to a
maximally entangled state.

Note that the recurrence method alone does not lead to a non-zero
rate since in every round we destroy or discard at least half of
the resources (all the target states) and maximally entangled
states are only obtained in the limit of infinitely many rounds.
To come to a rate, the resulting states after a finite number of recurrence steps are distilled by
the hashing/breeding protocol.

\subsection{The breeding protocol}
The breeding protocol is an entanglement assisted distillation
protocol adapted to Bell diagonal sates. In addition to the
state $\rho_\lambda$ Alice and Bob share arbitrarily many
maximally entangled states, which they can use during the
distillation process. At the end of the protocol they have to give
back the maximally entangled states they abused during the protocol.

Assume that Alice and Bob share $n$ copies of a Bell diagonal
state $\rho_\lambda$
 \beq\label{thestate}
\rho_\lambda^{\otimes n}=\sum_{k_1 \ldots k_n, l_1 \ldots l_n}
\lambda_{k_1l_1}\cdots\lambda_{k_n l_n} P_{k_1 l_1}\otimes
\cdots\otimes P_{k_n l_n}. \label{many}\eeq
 An appropriate interpretation of
Eq.(\ref{thestate}) is to say that Alice and Bob share the state
\beq P_{k_1 l_1}\otimes \cdots\otimes P_{k_n l_n}:=P_{\vec S} \eeq
with probability $\lambda_{k_1l_1}\cdots\lambda_{k_n l_n}$. Such a
sequence of  Bell-states can be identified with the string
\beq
\vec S=(k_1,l_1, \dots ,k_n,l_n). \eeq
%\beq
%\psi_{00}\otimes \psi_{10}\otimes \dots\psi_{00}\approx 0010\dots00
%\eeq
Note that if Alice and Bob knew
the sequence $\vec S$, they could
apply appropriate local unitary operations in order to obtain the
standard maximally entangled state $P_{00}^{\otimes n}$ and thus
gain $n$ ebits of entanglement.

It was shown in \cite{BennettPRL, BennettPRA} that given such a
string of Bell states and one extra maximally entangled state
$P_{00}$ one can check an arbitrarily parity of the string $\vec
S$, without changing or disturbing the sequence of Bell states.
This is done by applying a sequence of bilateral CNOT operations
with the extra maximally entangled state acting as target states,
and at a time one state of $\vec S$ as source state.
 The sequence of Bell states is unchanged, but the
maximally entangled state $P_{00}$ changed to $P_{x0}$, where $x$
can be any parity check of the vector $\vec S$, i.a. x can be
chosen to be \beq x=\sum_i S_i M_i=\braket{\vec S}{\vec
M},\label{l1} \eeq where $\vec M$ is an arbitrarily vector with
vector components  $\in {0,1}$. Note that Eq. (\ref{l1}) has to be
read modulo 2. By a local Bell measurement of the  target state
 Alice and Bob can gain the
information $x$, destroying the extra entangled state. For more
details see \cite{BennettPRA}. To such a single operation we will
refer as {\it parity check $\vec M$}. Remember that each such
parity check cost Alice and Bob one ebit, i.a. one copy from their
pool of maximally entangled states.

The main idea of the breeding protocol is to repeat such parity checks until the full sequence
$\vec S$ is known. If Alice and Bob did need $m$ parity checks to identify a sequence of $2n$ bits
(n qubits), then they gain $n-m$ maximally entangled states, giving them a distillation rate per copy of $1-m/n$.

This identifying process of $\vec S$ is done in the limit of the
numbers of copies $n$ going to infinity. This gives the advantages
to restrict in the identifying process to so called {\it typically
codewords} $\vec S$ \cite{Schumacher}. Note that at this stage the
whole distillation process is translated into a classical problem:
Given an classical bit string $\vec S$ generated by a probability
distribution $(\lambda_{ij})$, how many parity checks are
necessary to identify the bit string ?
 The String $\vec S$ generated
by many copies of the state $\rho_\lambda$ contains $n S(\rho_\lambda)$ bits of
classical information, where \beq S(\rho_\lambda)=-\sum_{ij}\lambda_{ij} \log
\lambda_{ij} \eeq is the von Neumann Entropy of the state $\rho_\lambda$. With
each parity check we gain one bit of this information \cite{proven}. Per qubit we
therefore need  $S(\rho)$ ebit of entanglement to identify it, leading to the well
known hashing/breeding rate \beq
D_{Hashqubit}(\rho_\lambda)=1-S(\rho_\lambda)=1-S(\lambda). \eeq The
hashing/breeding rate has shown to be optimal for rank deficient Bell states,
i.e., Bell diagonal states of rank two\cite{REbound2}.

The same kind of protocol can be adapted to states that are
diagonal in a tensor product basis of Bell states, i.e. for states
that are diagonal in a basis of the form
\beq \psi_{\vec i \vec
j}=\psi_{i_1j_1}\otimes \psi_{i_2j_2} \dots
\psi_{i_mj_m}\label{nbasis}, \eeq
i.e. states of the form
$\sigma_\lambda=\sum_{\vec i \vec j} \lambda_{\vec i \vec j}
P_{\vec i \vec j}$. In this case one copy of a state generates not
a two bit (one qubit) string, but a $2m$ bit (n qubit) string.
Taking $n$ copies of the state we get a random $2nm$ bit string
generated by the probability distribution $(\lambda_{\vec i \vec
j})$. We can  try to identify this string by doing parity checks
in the same way we did before. The amount of parity checks per
copy is in the same spirit given by $S(\sigma_\lambda)$, but if we
identified one copy of the state we now gain $m$ ebits instead of
1 ebit. The distillation rate for these states is therefore given
by \beq D_{Hash}(\sigma_\lambda)=m-S(\lambda).\label{haschn} \eeq
For more details see \cite{VW}. A special type of states that are
diagonal in such a basis are many copies of Bell diagonal states,
e.g.
 $\sigma_\lambda=\rho_\lambda^{\otimes n}$. But since the hashing/breeding rate (\ref{haschn}) is additive we gain or lose  nothing by applying
a breeding protocol to several copies of a Bell state.

\section{The new Protocol}\label{new}
The new protocol is essentially an asymptotic version of a recurrence step
followed by the breeding protocol.  The new idea is to take many copies of the
state $\rho$ and start by distilling states  which have the 'same parity'.
\subsection{Distill parity states}\label{parity}
In the breeding protocol the goal is to get the full information "$ij$" for every state $\psi_{ij}$.
One can also set the goal to get only a part of this information, e.g. we want only to know the $i$ or
the $j$ or the parity $i+j$. This correspond to the the information we would get, if we made the
corresponding parity check $\braket{\{i,j\}}{\vec m}$ on one copy, e.g. $\vec m=10,01,11$, which would cost  one ebit per copy.

But this information can be obtained in a much cheaper way using exactly the same asymptotic technics as in the
breeding protocol.
We can write the state as
\beq
\rho_\lambda=\mu_0 \rho_0 +\mu_1 \rho_1,
\eeq
 where
\bea
\mu_{k}&=&\sum_{\braket{\{i,j\}}{\vec m}=k}\lambda_{ij}\\
\rho_{k}&=&\frac{1}{\mu_{k}}\sum_{\braket{\{i,j\}}{\vec m}={k}}\lambda_{ij}P_{ij}.
\eea
If we take now $n$ copies of the state $\rho_\lambda$ we get
 \beq\label{gammel}
\rho_\lambda^{\otimes n}=\sum_{k_1 \ldots k_n}
\mu_{k_1}\cdots\mu_{k_n } \rho_{k_1}\otimes
\cdots\otimes \rho_{k_n}. \eeq
Similarly to (\ref{thestate}) an  appropriate interpretation of
Eq. (\ref{gammel}) is to say that Alice and Bob share the
state
\beq
\rho_{k_1}\otimes \cdots\otimes \rho_{k_n}, \eeq
which can be identified with a bit string
\beq
\vec S'=(k_1, \dots ,k_n).
\eeq
To get the information about $\vec S'$ in the same way we get it for the string $\vec S$, we need
 to make arbitrary parity checks $\vec M'$ on the vector $\vec S'$. Fortunately, $\vec S'$ and $\vec S$ described
exactly the same sequence of Bell state, the only difference is that $\vec S'$ contains not the full information.
Thus we can translate any parity check $\vec M'$ on $\vec S'$ to a parity check $\vec M$ on $\vec S$
by the simple rule
\beq
M_{2i-1}=m_1, M_{2i}=m_2
\eeq
whenever $M'_i=1$ and
\beq
M_{2i-1}=0, M_{2i}=0
\eeq
whenever $M'_i=0$.
It is readily checked that we can make this way every parity check
$\vec M'$ on $\vec S'$ paying on ebit of entanglement. To identify $\vec S'$ in the limit of many copies
we therefore need $S(\mu_0,\mu_1)$ ebit of entanglement.

So given a state $\rho$, we can decompose the state into $\rho_0$ and $\rho_1$ with probability
$\mu_0,\mu_1$ by paying $S(\mu_0,\mu_1)$ ebit per copy. Doing such a step in
our protocol we will refer as an {\it asymptotic parity check $\vec m$ }.
In a completely analog way we define asymptotic parity checks for
states that are diagonal in the basis (\ref{nbasis}). The parity vector $\vec m$ is in this case
a $2 m$ bit string. The rule to translate parity checks is given by
\beq
M_{(m-1) i+1, \dots, m i}=\vec m
\eeq
whenever $M'_i=1$ and
\beq
M_{(m-1) i+1, \dots, m i}=\vec 0
\eeq
whenever $M'_i=0$. $\vec 0$ denotes a vector of length $m$ containing only zeros.

If Alice and Bob make enough asymptotic parity checks, they gain the full information about
every copy, making this procedure equivalent to a distillation.
 Indeed, the original breeding protocol consist of two of such asymptotic parity checks.
First a  '$10$' asymptotic parity check and afterwards
a '$01$' asymptotic parity check.
 Since for the entropy holds
\bea
S(\lambda)&=&(\lambda_0+\dots+\lambda_m)S([\lambda_0,\dots,\lambda_m])\\
&+&(\lambda_{m+1}+\dots+\lambda_n)S([\lambda_{m+1},\dots,\lambda_n])\\
&+&S(\lambda_{1}+\dots+\lambda_{m},\lambda_{m+1}+\dots+\lambda_{n}), \eea we gain or lose
nothing by doing the distillation in several asymptotic parity
checking steps. Here $S([p_1,\dots,p_n])$ denotes the entropy of
the normalized probability distribution, i.e.
$S([p_1,\dots,p_n]):=S(p_1/N,\dots,p_n/N)$ with $N=\sum_i p_i$.

\subsection{The improved protocols}
The key of our new distillation protocols is to  distill several copies of a Bell diagonal state $\sigma_\lambda=\rho^{\otimes m}_\lambda$,
by a sequence of asymptotic parity checks.
After each such step, we get two kind of states, the one that pass the asymptotic parity check (parity equal zero)
and the one that fail the test (parity equal one).
Instead of continuing the  distillation by testing further parities, we consider two
new possibilities,
\begin{itemize}

 \item  We can decide to drop some of the states, dependent on the outcome of an asymptotic
        parity check.
         Indeed, this improves our distillation
        rate, iff the state we drop has a negative breeding rate, i.e. $m-S(\rho)<0$.
  \item We make a local Bell measurement on one of the $m$ two qubit systems. Dependent on the measurement output we get a
        new  $m-1$ two qubit system state. Before doing the measurement, we also allow to apply Alice and Bob
        local unitaries. But we restrict to such unitaries, that map many copies of Bell states again
        to many copies of Bell states.
\end{itemize}
\subsubsection{Asymptotic recurrence}
First we present a protocol, that is very similar to making one recurrence step and afterwards distilling
the state with the breeding protocol. The main difference is, that the recurrence step is somehow made
in an asymptotic way.
We take two copies of a Bell diagonal $\rho_\lambda$. Then we asymptotically check the parity $1010$ of
these two copies.
The state  passes this test, if it is in one of the states
$$0000,
0001,
0100,
0101,
1010,
1011,
1110,
1111,
$$
it fails the test if it is in one of the states$$0010,0011,0110,0111,1000,1001,1100,1101.$$
Therefore,  $\rho_\lambda^{\otimes 2}$ passes the parity check with probability
$p_{even}=(\lambda_{00}+\lambda_{01})^2+(\lambda_{10}+\lambda_{11})^2$ and is afterwards in the state
\beq
\rho_{even}=\frac{1}{p_{even}}\sum_{ijl} \lambda_{ij}\lambda_{il} P_{ij}\otimes P_{il}.
\eeq
With probability $p_{odd}=2(\lambda_{00}+\lambda_{01})(\lambda_{10}+\lambda_{11})$ it fails and is is in the state
\beq
\rho_{odd}=\frac{1}{p_{odd}}\sum_{ijl} \lambda_{ij}\lambda_{(i+1)l} P_{ij}\otimes P_{(i+1)l}
\eeq
This step costs  $S(p_{odd},p_{even})$ ebit per copy of $\rho_\lambda^{\otimes 2}$. Similarly to the recurrence protocol,
we decide to drop one of these two possible outcomes.
A calculation for Werner states shows, that $\rho_{odd}$ has a negative breeding rate for
every entangled Werner states $\rho_f$. So we decide to drop all odd states and continue to distill the even states. The distillation
rate is then given by
\beq
-S(p_{odd},p_{even})+p_{even}(2-S(\rho_{even})).
\eeq
$S(p_{odd},p_{even})$ is what we have to pay for the first parity check. Afterwards we continue with
probability $p_{even}$ to distill $\rho_{even}$. Normalized to one copy of the state $\rho_\lambda$ we obtain
\beq
-S(p_{odd},p_{even})/2+p_{even}(1-S(\rho_{even})/2),
\eeq
which is plotted in Fig.\ref{fig1} for Werner states. Surprisingly, we get an improvement to the known
breeding protocol over the full range of the fidelity and not only for low fidelities, like it is
known for combined hashing recurrence protocols.

\subsubsection{ the 2-copy rate}
\begin{figure}\epsfig{file=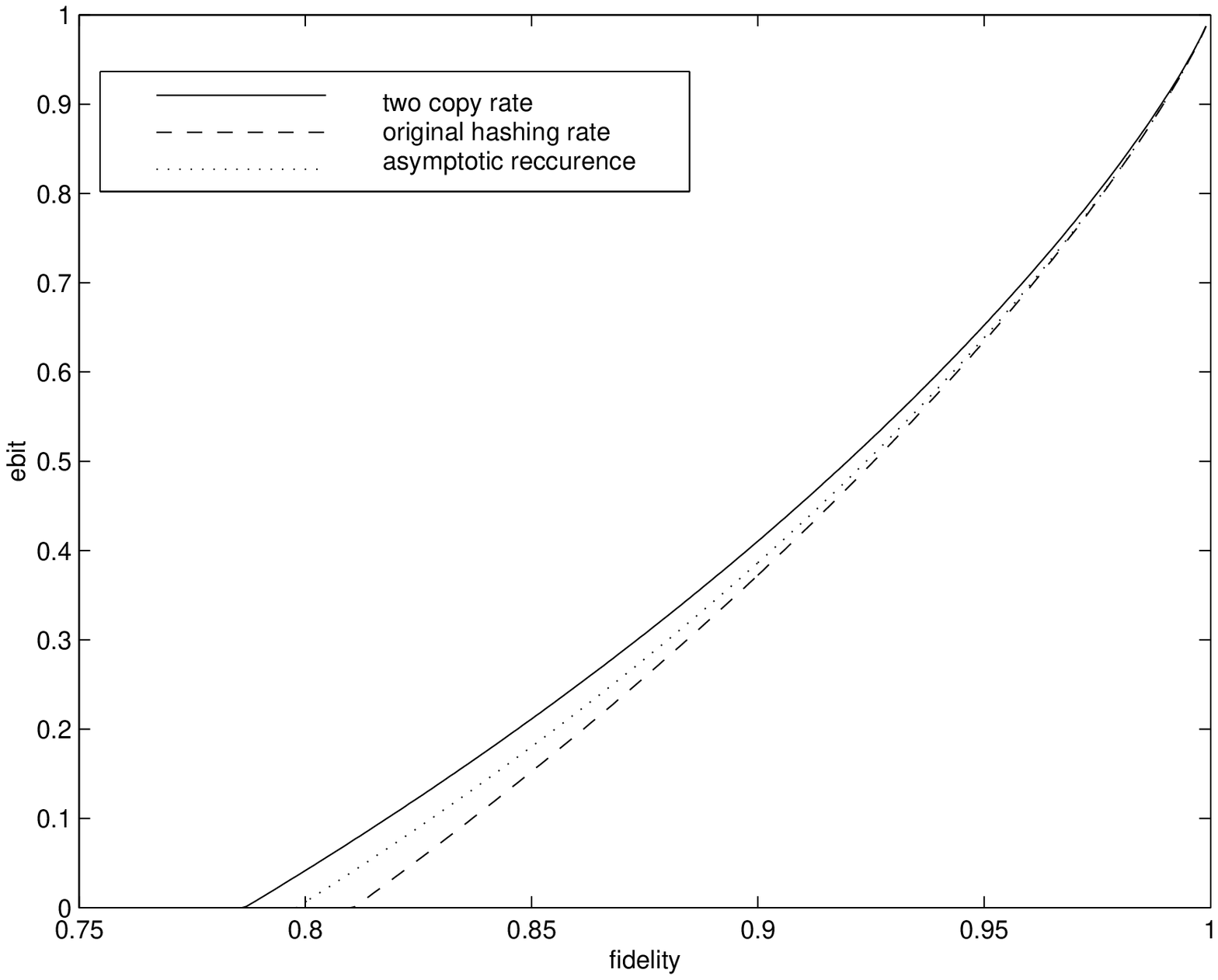,width=9cm}
\caption{\label{fig1}Distillation rates for Werner states.}
\end{figure}

Although the breeding rate of the state $\rho_{odd}$ is negative, the state may be still entangled and even distillable.
So one can improve the distillation rate, by applying an alternative distillation protocol onto $\rho_{odd}$.
One way to get a positive rate out of $\rho_{odd}$ is to make a local Bell measurement on one of  the two qubit systems.
If the measurement outcomes of Alice and Bob do not coincide (correspond to the states $10$ or $11$), they knew that the remaining
state has to be
$00$ or $01$, because they knew, the overall state had an odd parity in the first bit of each two qubit system. If the
measurement outcomes coincide ($00$ or 01), they knew the remaining state in one of the states $10,11$. So for both outcomes they end up
in a rank two Bell diagonal state, for which they knew, that the breeding protocol gives them the optimal distillation rate.
In fact, the mixture of $00,01$ is always equally weighted and therefore  separable, so they can drop it. But the mixture of $10,11$ can give
positive contribution to the overall distillation rate. The distillation rate obtained with this protocol
for Werner states is plotted in Fig.\ref{fig1}.

The distillation rate for arbitrary Bell diagonal two qubit states using this protocol is easily calculated to be
\bea
D_{2c}(\vec\lambda)&=&1-S(\rho)\\&&+ \frac{p_{odd}}{4}
[S([\lambda_{00},\lambda_{01}])+S([\lambda_{11},\lambda_{10}]) ] \label{extra}
\eea
 So it is a real improvement to the known hashing/breeding rate, since we beat this
 rate by the extra term (\ref{extra}), which is always larger or equal to zero. Indeed,
 (\ref{extra}) is zero if and only if the state  has rank  less than two.
We have a strictly positive improvement for all Bell diagonal states, with the only exception for
those cases, where the hashing/breeding rate was known to be optimal.

\subsection{Further optimization}

What rates can we achieve using protocols consisting only out of asymptotic
parity checks and local Bell measurements ?

\begin{figure}\epsfig{file=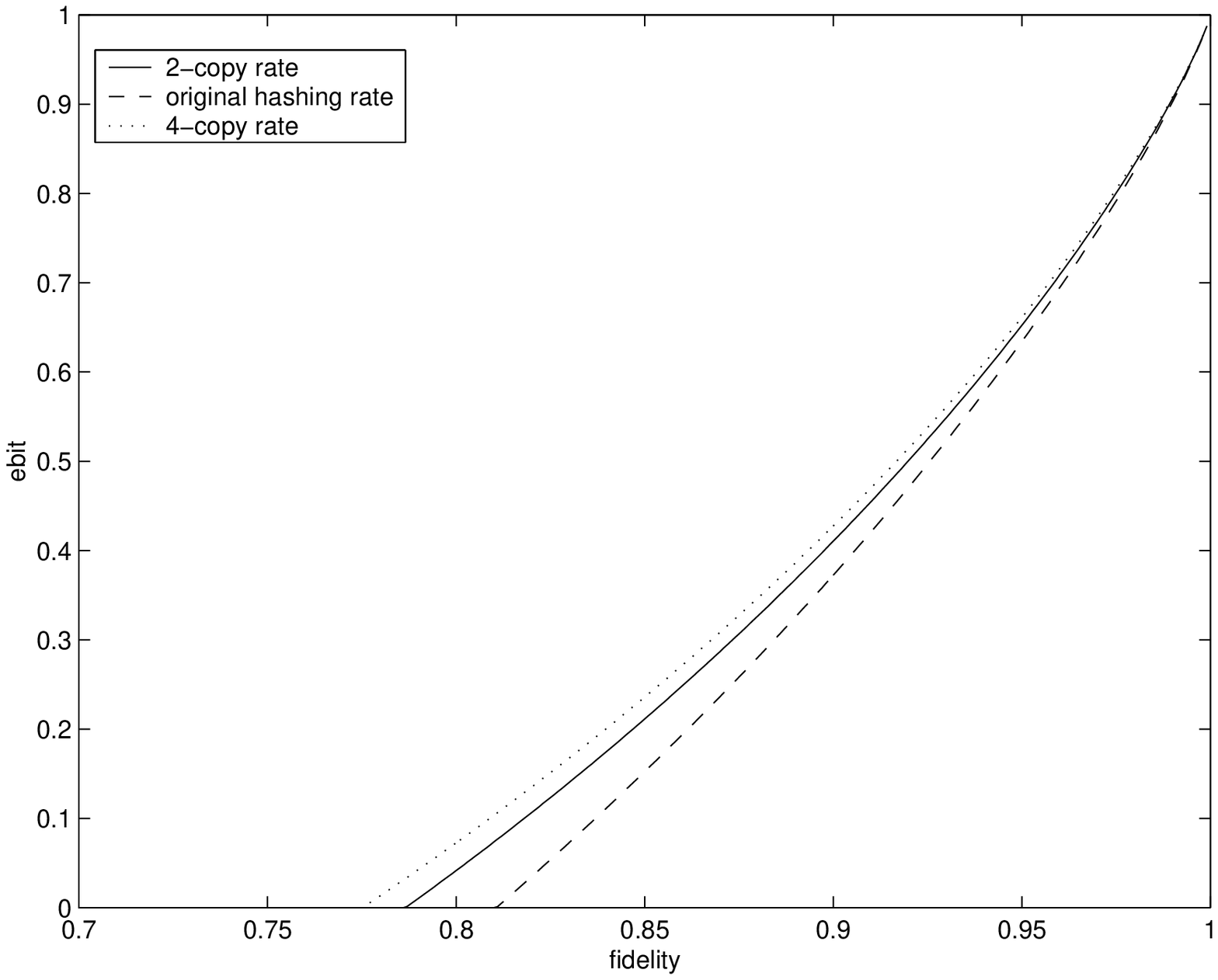,width=9cm}
\caption{\label{fig4c}Distillation rates for Werner states.}
\end{figure}

We can start from more than two copies of the state and construct in the same
spirit protocols consisting out of asymptotic parity checks and local Bell
measurements. The result of one such protocol starting with four copies
$\rho_\lambda^{\otimes 4}$ is plotted in Fig.\ref{fig4c} for Werner states.
Using self-recurrent programming, it is in principle possible to prove optimality
of such schemes within the class of operations described above. Note however that
the computational cost for doing this optimization is superexponential in the
number of copies involved in the protocol, and proving optimality therefore does
not seem to be feasible for more than 6 copies. It is however clear that taking
more copies can only increase the distillation rate. Unfortunately, we did not
manage to find an easy way to construct reasonable good protocols for more and more
copies; in the asymptotic limit, this would maybe even lead to  an exact
expression for the entanglement of distillation.

On Werner states the above 2-copy protocol is essentially the best
possible protocol using asymptotic hashing steps and local
measurements. We have done an optimization  for Werner states over
all protocols, which include arbitrary asymptotic hashing steps,
local Bell measurements and arbitrary local unitary operations,
mapping multi-copies of Bell states into multi-copies Bell states.
The number of such protocols is finite. The Asymptotic hashing
steps can be parameterized by the $2n$ bit string $\vec m$. All
local unitaries that map multi-copies of Bell states to
multi-copies of Bell states have been characterized in
\cite{Frank}. For two copies of Bell states there exists up to
local equivalence 120 different possibilities. The number of
possible asymptotic parity checks or local Bell measurements  that
are possible, until the state is known or completely destroyed is
also finite.

\begin{figure}\epsfig{file=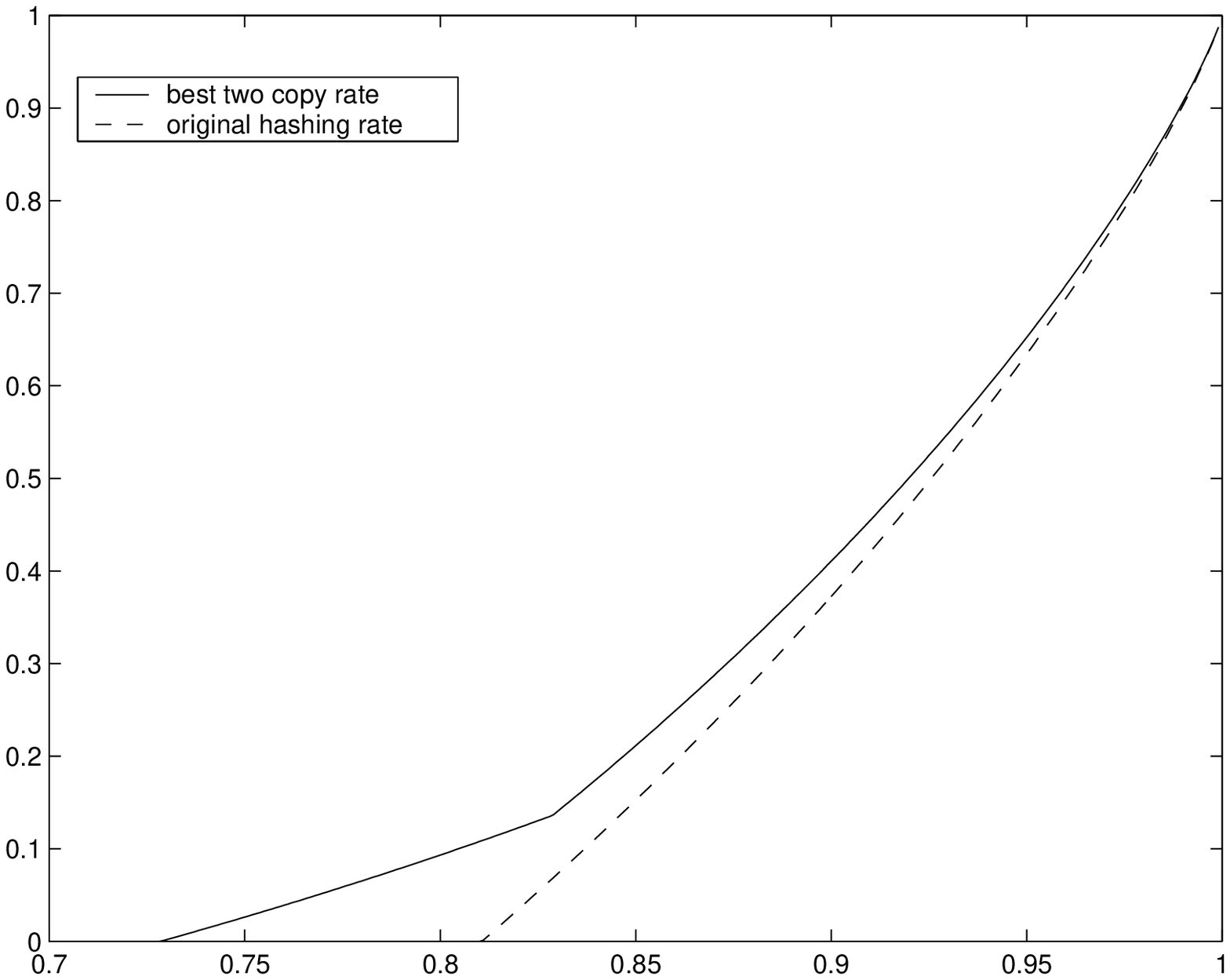,width=9cm}
\caption{\label{fig2}Distillation rates for Werner states.}
\end{figure}

So one has to test for a given Bell diagonal state a finite (but very large) set
of possible protocols. The result of this optimization for all protocols starting
from two copies of a   Werner  state is plotted in Fig.\ref{fig2}. For high
fidelities the above 2-copy protocol gave the best results. For low fidelities the
standard recurrence followed by the optimized hashing, which is included in this
kind of protocols, is optimal.

For more than two copies of the state we were not able to do the full optimization
over all possible protocols, because the number of protocols is too large. By
making a good guess, we could find protocols up to 5 copies giving further small
improvements. But so far, we are unable to give a systematic way of producing
n-copy-protocols.

\section{Non Entanglement assisted protocols}
We will now show, how the above protocols can be transformed into  protocols that
do not need any pre-distilled maximally entangled states. The idea is to start
with an non entanglement assisted protocol to get a starting pool of maximally
entangled states for the entanglement assisted protocol. To distill this starting
pool, we will only use a sub-linear amount of copies, so that in the asymptotic
limit, we will not affect the obtained rates.

In detail, we start with $k^{-1}\sqrt{n}+n$ copies of the state $\rho$, where
$k>0$ is the distillation rate obtained by any non entanglement assisted protocol,
e.g. we can use several recurrence steps followed by the hashing  protocol. We use
this non entanglement assisted {\it activating  protocol} to distill $\sqrt{n}$
maximally entangled states out of the first $k^{-1}\sqrt{n}$ copies. At this step
let us assume, that the activating protocol does  this without any error, i.e. we
get perfect maximally entangled states.

We use these $\sqrt{n}$ maximally states as resource for our entanglement assisted
protocol to distill $(1+r)\sqrt{n}$ copies of the state $\rho$ gaining
$(1+r)\sqrt{n}$ new maximally entangled states. At this step we  also assume the
entanglement assisted protocol to work perfect for a finite amount of copies. Here
$r$ denotes the rate obtained by the entanglement assisted protocol. With these
maximally entangled states we distill the next block of $(1+r)^2 \sqrt{n}$ copies
from the $n$ copies of $\rho$ and  so on for all the rest of it. The obtained rate
for this non entanglement assisted protocol will be $r$, independent of the rate
$k$ of the activating protocol at the beginning, because the extra
$k^{-1}\sqrt{n}$ copies of $\rho$ will be vanish compared to $n$ for $n$ going to
infinity.

Of course, in contradiction to our assumption, none of the protocols will work
perfect for any finite number of copies  and the errors, that will occur, may
accumulate and destroy the whole protocol. So to  claim this above protocol to be
a valid distillation protocol, we have to ensure that the success probability of
the protocol goes to one if $n$ goes to infinity.

So we now want to bound the success probability of the protocol described above.
First of all we only can guarantee our protocol work, if the starting $\sqrt{n}$
maximally entangled states were correct, i.e. if the activating protocol
succeeded. Let us call this probability $p(n)$, with $p(n)$ going to zero, if n
goes to infinity. Second we call $q(n)$ the probability, that the entanglement
assisted protocol works well. Then the probability that the above protocol works
can be bounded by
$$p_{succ}\geq p(\sqrt{n}) q(\sqrt{n})^{\sqrt{n}}.$$
In the worst case we have to repeat the entanglement assisted protocol $\sqrt{n}$
times successfully with only $\sqrt{n}$ copies of the state $\rho$ as input.
Indeed, this probability can go to zero, if $q(n)$ scales in a wrong way. The
probability $p(\sqrt{n})$ is no problem at all, as long as it goes to one for
large $n$.

So we only have to ensure, that $q(n')^{n'}$ with $n'=\sqrt{n}$ goes to one as
$n'$ goes to infinity. What we therefore need is an approximation of the scaling
of the success probability $q(n)$ for our new protocols. The entanglement assisted
protocol described above will succeed, if the sequences correspond to so called
typical sequences. The probability of being a typical sequence goes exponentially
to one \cite{Frankcopy}, so we can bound for $n'$ large enough $q(n')> (1-k e^{-c
n'})$. So the overall success probability can be bounded by
$$p_{succ}\geq p(\sqrt{n})(1-k e^{-c\sqrt{ n}})^{\sqrt{n}},$$
which goes to one, if $n$ goes to infinity.

\section{Conclusion}
We have presented the first distillation protocol, where two way communication is
included in an asymptotical sense. The distillation rates attained this way beat
the known hashing/breeding rate over the whole range of entangled Bell diagonal
states. Since the hashing/breeding protocol is the last step for all recurrence
like distillation schemes, we can improve all these distillation protocols.
Furthermore we proved how any entanglement-assisted distillation protocol can be
converted into a non-entanglement-assisted distillation protocol with the same
rate. An important open problem is to find a systematic way for constructing
n-copy distillation protocols; this could ultimately lead to an exact expression
of the entanglement of distillation.

Work supported by EU IST projects, the DFG, and the
Kompetenz\-netz\-werk Quanten\-informations\-verarbeitung der
Bayerischen Staatsregierung.


\begin{references}
\bibitem{proven} Indeed, to show that we gain 1 bit of information and not less is the tricky part in the proof.
See \cite{BennettPRA}.
\end{references}
\end{document}